\documentclass[]{aastex7}
\usepackage{graphicx}
\usepackage{fancybox}
\usepackage{ascmac}
\usepackage{amsmath}
\usepackage{amssymb}
\usepackage{bm}
\usepackage{amsmath}

\begin{document}

\title{Dynamical Evolution of V-Shaped Collision Debris}

\author[orcid=0000-0003-4590-0988]{Ryuki Hyodo}
\affiliation{Earth-Life Science Institute, Institute of Science Tokyo, 2-12-1 Ookayama, Meguro-ku, Tokyo, 152-8550, Japan}
\affiliation{Universit\'{e} Paris Cit\'{e}, Institut de Physique du Globe de Paris, CNRS, F-75005 Paris, France}
\affiliation{Graduate School of Artificial Intelligence and Science, Rikkyo University, 3-34-1 Nishi-Ikebukuro, Toshima-ku, Tokyo 171-8501, Japan}
\affiliation{SpaceData Inc., 1-17-1 Toranomon, Minato-ku, Tokyo 105-6490, Japan}
\email{ryuki.h0525@gmail.com}

\author[orcid=0009-0003-5452-7473]{Naoya Torii}
\affiliation{Earth-Life Science Institute, Institute of Science Tokyo, 2-12-1 Ookayama, Meguro-ku, Tokyo, 152-8550, Japan}
\email{torii@elsi.jp}


\begin{abstract}
Catastrophic collisions between proto-satellites have been proposed as a possible origin of Saturn's rings. This argument relies on the concept of the equivalent circular orbit. Here, we re-examine the post-impact dynamical evolution of collision debris using analytical arguments and $N$-body simulations with fragmentation. We focus on the long-term evolution of debris distributed in a broad V-shaped region in the $a$--$e$ plane, with two arms for particles sharing a common collision radius. Because particles on the two arms possess significantly different angular momenta, inter-arm collisions dominate the evolution and drive behavior fundamentally different from the simple circularization assumed in the equivalent circular orbit approach. As a result, the classical equivalent circular orbit concept cannot predict the long-term fate of collision debris. Both our analytical framework and $N$-body simulations show that, although some debris initially passes within the Roche limit on eccentric orbits, successive collisional evolution drives the particles approximately along the original V-shaped constraint curves toward the apex of the V-shape, i.e., the original collision radius. Instead of spreading inward to form a ring, the debris converges and reaccretes near the original collision location. We therefore conclude that catastrophic proto-satellite collisions do not produce massive Saturnian rings. Rather, the debris evolves toward reaccretion into a new generation of satellite-sized bodies near the impact radius. These results fundamentally revise the dynamical interpretation of collision-generated debris and establish a more general framework applicable beyond the Saturnian system, including other planetary ring systems and debris produced during planet formation.
\end{abstract}



\section{Introduction}
\label{sec_introduction}

\setcounter{footnote}{0} 

Saturn's main rings, composed predominantly of water ice,
have long been considered a primordial feature of the Saturnian system.
However, recent analyses of Cassini data have fundamentally challenged this view.
The low mass of the rings measured by Cassini \citep{Ies19},
combined with constraints from the micrometeoroid bombardment flux, which could pollute the rings over time \citep{Kem23}
and viscous evolution modeling \citep{Est23},
consistently point to a ring exposure age of no more than a few hundred million years
\citep[see also][for a review]{Cri25}.
However, \citet{Hyo25} showed that micrometeoroid bombardment
does not necessarily lead to pollution of the rings:
impacting micrometeoroids can vaporize upon collision with ring particles,
and the resulting vapor may be lost through recondensation into fine grains
and subsequent interaction with Saturn's magnetosphere,
rather than being simply deposited onto ring particles
as previously assumed.
This implies that the high ice purity of the rings
does not necessarily require a young age,
and that the rings could potentially be as old as the solar system \citep[see also][]{Cri19}.

Several mechanisms have been proposed for the formation of Saturn's rings,
and they can be broadly divided into two categories
depending on the epoch at which the rings are formed.
In the first category, the rings are primordial:
they formed, for example, during the early stages of solar system evolution
through the tidal disruption of a large differentiate body,
such as a Kuiper belt object or a comet,
passing within the Roche limit of Saturn \citep{Don91,Hyo17c} or the tidal disruption of a primordial circumplanetary large moon that spiraled inward through Saturn's circumplanetary gas disk \citep{Can10}\footnote{The Roche limit,
inside which tidal forces from the planet exceed
the self-gravity of an orbiting body, is given by
$a_{\rm Roche} = 2.456 \, R_\mathrm{planet}
(\rho_\mathrm{planet}/\rho_\mathrm{particle})^{1/3}$,
where $R_\mathrm{planet}$ and $\rho_\mathrm{planet}$ are the planetary radius
and mean density, and $\rho_\mathrm{particle}$ is the bulk density of the orbiting body.
For Saturn with icy particles ($\rho_\mathrm{particle} = 900\;\mathrm{kg\,m^{-3}}$),
$a_{\rm Roche} \approx 2.25\,R_\mathrm{planet}$}.
In the second category, the rings are young:
\citet{Cuk16}, for example, proposed that a dynamical instability among Saturn's satellite system
led to catastrophic collisions between mid-sized moons
within the last few hundred million years,
and that the debris produced by such a recent collision
spread inward to populate the Roche zone,
forming the observed ring system\footnote{Alternatively, the tidal disruption of a pre-existing moon after orbital destabilization and a close encounter with Saturn is proposed, as a scenario for the formation of young rings \citep{Wis22}. However, a detailed comparison with these alternative ring-formation scenarios is beyond the scope of this study.}.

Following the scenario proposed by \citet{Cuk16},
\citet{Hyo17d} performed smoothed particle hydrodynamics (SPH) simulations
of a catastrophic collision between two Rhea-sized satellites orbiting Saturn,
and subsequently followed the dynamical evolution of the resulting debris
using $N$-body simulations.
They demonstrated that at every stage of the debris evolution,
the timescale for mutual accretion among debris particles
is shorter than the timescale for viscous spreading of the debris disk.
As a result, the debris rapidly reaccretes into large satellite-sized bodies
rather than spreading inward toward the Roche limit.
This finding suggested that, contrary to the hypothesis of \citet{Cuk16},
a satellite collision cannot produce Saturn's rings.

More recently, \citet{Teo23} and \citet{Cuk26} revisited the satellite collision scenario
using SPH simulations with significantly higher resolution than those of \citet{Hyo17d}.
With this increased resolution,
they showed that the collision debris is distributed over a wider range
of semi-major axes and eccentricities in orbital element space.
They then introduced the ``equivalent circular orbit''
(also referred to as the ``equivalent circular radius'')
to evaluate whether this debris can \textit{directly} reach the Roche limit.
For a particle on a Keplerian orbit with semi-major axis $a$,
eccentricity $e$, and inclination $i$ relative to the equatorial plane,
the vertical component of the specific angular momentum is
$L_{\rm z} = \sqrt{\mu a(1-e^2)}\cos i$, where $\mu = GM$ is the gravitational parameter
of the central planet.
The equivalent circular radius $a_\mathrm{eq}$ is defined as the radius of the circular,
equatorial orbit that possesses the same angular momentum \citep{Can04}:
\begin{equation}
  a_\mathrm{eq} = a(1-e^2)\cos^2 i.
  \label{eq:aeq}
\end{equation}

\citet{Teo23} and \citet{Cuk26} applied this concept
to the orbital elements of the debris immediately after the SPH collision,
and argued that particles satisfying $a_\mathrm{eq} < a_{\rm Roche}$
will eventually circularize within the Roche limit $a_{\rm Roche}$---the region
where Saturn's rings exist today.
They thereby concluded that a sufficient mass
to account for the present-day ring mass of Saturn
can be delivered as circularized orbits inside the Roche limit
through a catastrophic collision between mid-sized moons.

However, there is a critical problem in their estimate.
They did not directly study the long-term dynamical evolution
of the collision debris through, for example, $N$-body simulations
as done by \citet{Hyo17d}.
Instead, they relied solely on the equivalent circular radius
computed from the orbital elements of the debris immediately after the collision
to assess how much debris would end up inside the Roche limit implicitly assuming circularization at constant angular momentum.
In this paper, we use both analytical arguments
and $N$-body numerical simulations.
We show that, although the debris can reach within the Roche limit
on eccentric orbits immediately after the impact,
the equivalent circular orbit approach,
as applied by \citet{Teo23} and \citet{Cuk26},
is not appropriate for assessing the long-term fate of the post-collision debris.
Thus, we show that the collision debris cannot form rings,
even though they initially pass within the Roche limit
on their post-impact eccentric orbits. If the post-impact debris
do not pass within the Roche limit, the debris simply circularizes and reaccretes into large bodies, since the timescale for mutual accretion is shorter than
that for viscous spreading \citep{Hyo17d}. In either case, rings are not produced\footnote{Our study specifically addresses the ring-formation aspect of the scenarios proposed by \citet{Teo23} and \citet{Cuk26}. Other aspects of those scenarios are beyond the scope of this work and are not ruled out here.}.

Our paper is organized as follows.  In Section~\ref{sec_analytical}, we develop the analytical framework
for debris evolution in the $a$--$e$ diagram,
deriving the constraint curves, the critical coefficient of restitution,
and the conditions under which debris can reach the Roche limit.
In Section~\ref{sec_numerical}, we present numerical simulations
that confirm and extend the analytical results. We summarize our conclusions in Section~\ref{sec_summary}.\\

\begin{figure*}[tbp]
	\begin{center}
	\includegraphics[width=0.9\textwidth]{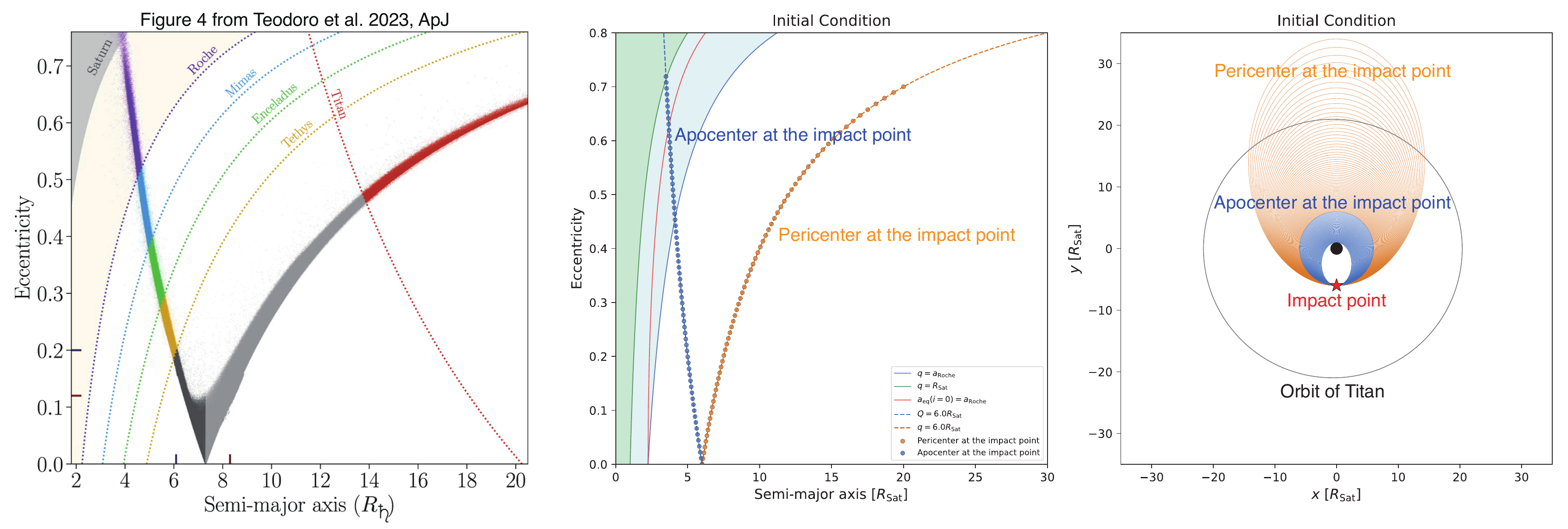}
	\caption{V-shaped distribution of collision debris in the $a$-$e$ diagram. Left: Post-impact debris from \citet{Teo23}. Middle: Initial V-shaped configuration adopted in this study, defined by the apoapsis ($a_{\rm col}=a(1+e)$) and periapsis ($a_{\rm col}=a(1-e)$) constraint curves. The blue and red curves mark the conditions $q=a_{\rm Roche}$ and $a_{\rm eq}=a_{\rm Roche}$, respectively, where $q=a(1-e)$ is the periapsis distance and $a_{\rm eq}$ is the equivalent circular radius. Green curves indicate $q=R_{\rm Sat}$. Right: Corresponding orbital trajectories in Saturn's equatorial plane, illustrating that particles on the two arms possess distinct angular momenta and orbital geometries, despite intersecting at the common radius $a_{\rm col}$.}
	\label{fig_schematic}
	\end{center}
\end{figure*}

\section{Analytical Approach}
\label{sec_analytical}

\subsection{Fundamental consideration of the post-impact dynamical evolution of debris}
The fundamental reason why the equivalent circular orbit approach
fails for post-collision debris is that the debris distribution
in the $a$--$e$ diagram has a characteristic V-shape
(Figure~\ref{fig_schematic}).
If the debris were distributed along only one arm of the V and if particles share similar angular momenta especially among adjacent particles, mutual collisions would gradually circularize their orbits
eventually toward radii close to the equivalent circular radius (see Appendix~\ref{appendix_twobody} and Appendix~\ref{appendix_one_arm}).
However, the V-shape consists of two distinct arms:
one arm corresponds to particles whose pericenters $q$ coincide with
the impact point $a_{\rm col}$ (orange points and lines in Fig.~\ref{fig_schematic}), and the other to particles whose apocenters $Q$
coincide with the impact point (blue points and lines in Fig.~\ref{fig_schematic}).
Particles belonging to different arms have very different angular momenta
and follow very different orbits (right panel of Fig.~\ref{fig_schematic}).

In a realistic catastrophic impact, the ejecta should generally possess a finite radial velocity dispersion at the release point. Therefore, the debris is not expected to lie exactly on the idealized curves $a_{\rm col} = a (1\pm e)$, but rather to occupy a finite-width region inside the V-shaped envelope defined by these two branches. In the present study, however, we first focus on the idealized V-shaped boundaries in order to isolate the fundamental dynamical mechanism governing the subsequent collisional evolution of broadly distributed impact debris.

Another important consideration in the case of a V-shape structure 
is the synodic period between particles on crossing orbits. 
For two particles with semimajor axes $a_1$ and $a_2$, 
their mean motions are $n_i=\sqrt{GM_{\rm Sat}/a_i^3}$, and the synodic period is given by\footnote{For reference, the orbital period at $6R_{\rm Sat}$ is $\sim 0.007$\,year.}
\begin{equation}
\label{eq:Tsyn}
T_{\rm syn} 
\simeq \frac{4\pi}{3}\frac{a}{n\,|\Delta a|}
= \frac{4\pi}{3}\frac{a^{5/2}}{\sqrt{GM_{\rm Sat}}}\frac{1}{|\Delta a|} \sim 0.03 \, \mathrm{yr} \left( \frac{a}{6R_{\rm Sat}} \right)^{5/2} \left( \frac{|\Delta a|}{R_{\rm Sat}} \right)^{-1},
\end{equation}
where $G$, $M_{\rm Sat}$, and $R_{\rm Sat}$ are the gravitational constant, mass of Saturn, and radius of Saturn, respectively.

Thus, the synodic period is inversely proportional to the 
difference in semimajor axis, $T_{\rm syn} \propto |\Delta a|^{-1}$. 
Collisions between particles belonging to different arms, 
which typically have larger effective $|\Delta a|$, 
therefore occur on a much shorter timescale than collisions 
between particles on the same arm. 
These inter-arm collisions govern the orbital evolution of the debris, 
leading to a dynamical outcome fundamentally different from 
the simple circularization assumed in the equivalent circular orbit approach.

We also consider another potentially important timescale, namely the apsidal precession timescale due to the planetary $J_2$ term \citep{Kau66}:
\begin{equation}
\label{eq:Tpre}
	T_{\rm pre} = \frac{2\pi}{|\dot{\varpi}|} 
	\sim 5 \, \mathrm{yr} \left(1 - e^{2}\right)^{2} \left(1 - \frac{5}{4}\sin^{2} i - \frac{1}{2}\cos i\right)^{-1} \left(\frac{J_{2}}{0.016298}\right)^{-1} \left(\frac{a}{6R_{\rm Sat}}\right)^{7/2} ,
\end{equation}
where $\varpi \equiv \Omega + \omega$ is the longitude of pericenter, $\Omega$ is the longitude of the ascending node, and $\omega$ is the argument of pericenter. Although the exact value depends on $e$ and $i$, we generally expect $T_{\rm syn} \ll T_{\rm pre}$. 

More directly, the relevant timescale for the $J_2$ term to destroy the initial apsidal alignment is not the absolute apsidal precession period, but the differential apsidal-precession period between neighboring orbits. Since Eq.~(\ref{eq:Tpre}) gives $\dot{\varpi}\propto a^{-7/2}$ for fixed $e$ and $i$, two nearby orbits separated by $|\Delta a|$ have, to leading order in $|\Delta a|/a$,
\begin{equation}
\label{eq:delta_dot_pi}
	|\Delta \dot{\varpi}|
	\simeq
	\frac{7}{2} |\dot{\varpi}| \frac{|\Delta a|}{a}.
\end{equation}
Thus, we define the apsidal synodic period, $T_{\rm syn,prec}$, as the time required for the longitudes of pericenter of two orbits separated by $|\Delta a|$ to differ by $2\pi$ owing to differential $J_2$-driven apsidal precession. It is then given by $T_{\rm syn,prec} \equiv 2 \pi / |\Delta \dot \varpi|$. Using Eqs.~(\ref{eq:Tsyn})--(\ref{eq:delta_dot_pi}), we obtain
\begin{equation}
	\frac{T_{\rm syn,prec}}{T_{\rm syn}}
	\sim
	290 
	\left(1 - e^{2}\right)^{2}
	\left(1 - \frac{5}{4}\sin^{2} i - \frac{1}{2}\cos i\right)^{-1}
	\left(\frac{J_{2}}{0.016298}\right)^{-1}
	\left(\frac{a}{6R_{\rm Sat}}\right)^{2} .
\end{equation}
This ratio is independent of $|\Delta a|$, because both $T_{\rm syn}$ and $T_{\rm syn,prec}$ scale as $|\Delta a|^{-1}$. Therefore, for the parameters considered here, the collisional synodic timescale is much shorter than the differential apsidal-precession timescale. In other words, collisions between particles on crossing orbits occur before differential $J_2$-driven apsidal precession can destroy the initial apsidal alignment of the debris cloud. Thus, differential apsidal precession does not dynamically decouple the two branches during the initial collisional phase. The effect of the planetary $J_2$ term is included in our $N$-body simulations. The numerical results below are consistent with this analytical expectation, showing that collisions dominate the orbital evolution before significant $J_2$-driven apsidal precession develops.

\begin{figure*}[htbp]
	\begin{center}
	\includegraphics[width=0.7\textwidth]{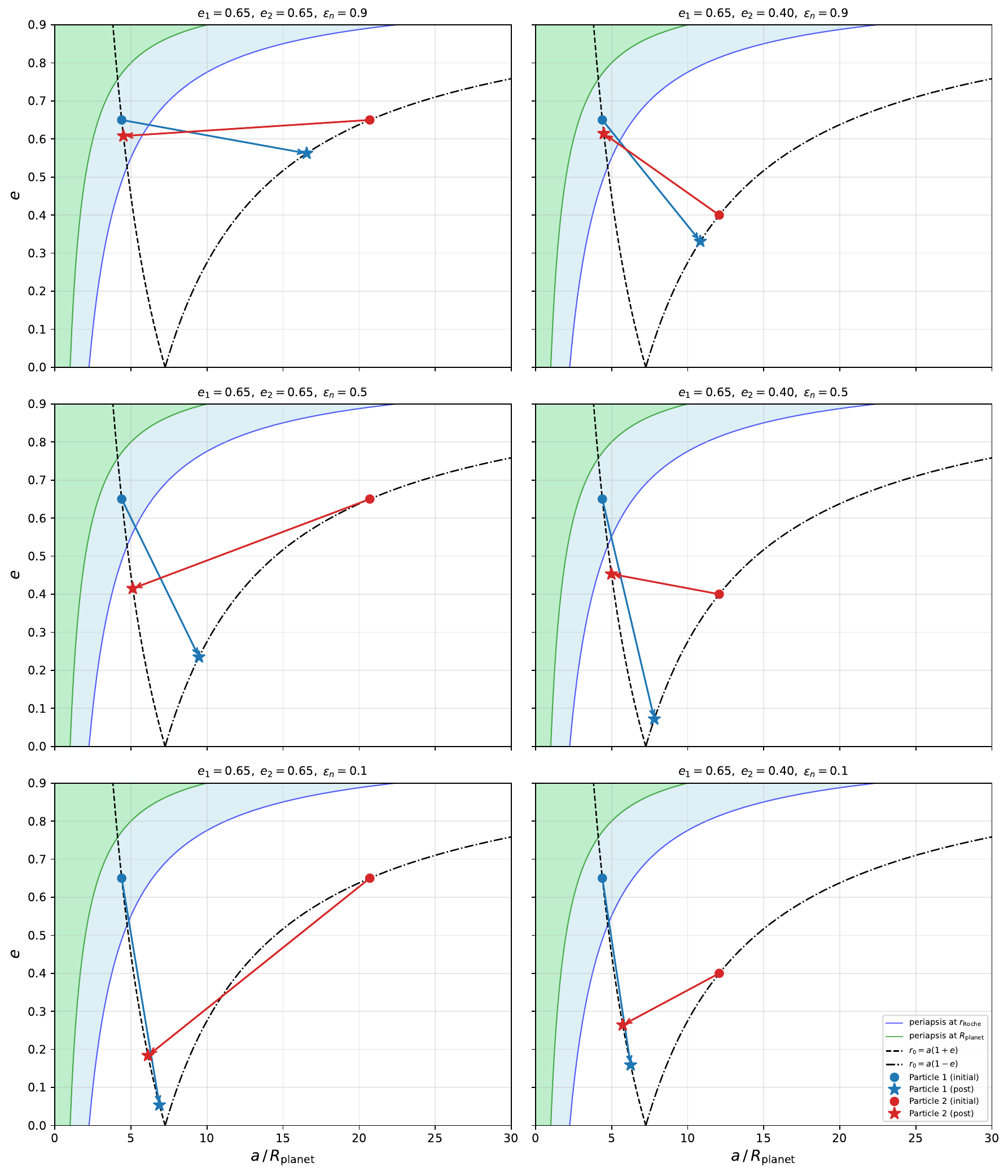}
	\caption{Post-collision orbital evolution of two particles in the $a$--$e$ diagram.
	Each panel shows the constraint curves for a collision at $a_{\rm col} = 7\,R_\mathrm{planet}$:
	the apoapsis constraint ($e = a_{\rm col}/a - 1$; left branch) and
	the periapsis constraint ($e = 1 - a_{\rm col}/a$; right branch).
	The green- and blue-shaded regions indicate orbits with periapsis
	below the planetary surface and the Roche limit, respectively.
	Blue and red symbols trace the successive positions of particle~1
	(initially at apoapsis) and particle~2 (initially at periapsis), respectively.
	The left column adopts $e_1 = e_2 = 0.65$,
	while the right column adopts $e_1 = 0.65$ and $e_2 = 0.40$.
	From top to bottom, the coefficient of restitution decreases:
	$\varepsilon_n = 0.9$, $0.5$, and $0.1$.
	For high $\varepsilon_n$, particle~1 crosses from the apoapsis curve
	to the periapsis curve, whereas for low $\varepsilon_n$,
	particles remain on their respective constraint curves
	and converge toward the circular orbit at $(a_{\rm col}, 0)$.}
	\label{Fig_evolution_analytical}
	\end{center}
\end{figure*}

\subsection{Fundamental dynamical evolution of V-shape distributed debris clouds}
Here, we develop an analytical framework to examine how such V-shaped post-collision debris clouds evolve in orbital element space. We consider two particles on intersecting Keplerian orbits that collide at a common radial distance $a_{\rm col}$ from Saturn, and derive the resulting changes in $a$ and $e$ before and after the collision. Figure~\ref{Fig_evolution_analytical} illustrates the evolution of the particles in the $a$-$e$ diagram, showing how the post-collision orbital elements remain confined to the original V-shaped constraint curves and evolve toward $(a,e) = (a_{\rm col},0)$ as orbital energy is dissipated. The detailed setup and analytical derivations are described below.

We consider two test particles orbiting Saturn
(mass $M$, radius $R_\mathrm{planet}$, mean density $\rho_\mathrm{planet}$)
in the same orbital plane. Each particle follows a Keplerian orbit characterized by the semi-major axis $a$ and eccentricity $e$. The two particles are arranged so that the apoapsis of particle~1 ($Q= a(1+e)$) coincides with the periapsis of particle~2 ($q = a(1-e)$) at a common distance $a_{\rm col}$ from the planet (i.e., $a_{\rm col}$ is the impact point):
\begin{equation}
  a_{\rm col} = a_1(1+e_1) = a_2(1-e_2).
  \label{eq:r0_constraint}
\end{equation}
Particle~1 initially lies on the apoapsis curve and particle~2 on the periapsis curve
(Figure~\ref{Fig_evolution_analytical}). These two curves meet at $(a, e) = (a_{\rm col}, 0)$, corresponding to a circular orbit at $a_{\rm col}$.

At any point on a Keplerian orbit, the orbital speed is given by $v = \sqrt{\mu\left(\frac{2}{r} - \frac{1}{a}\right)}$. At an apsis (periapsis or apoapsis), the radial velocity vanishes and the velocity is
purely tangential.
Therefore, at $r = a_{\rm col}$:
\begin{align}
  v_1 &= \sqrt{\mu\left(\frac{2}{a_{\rm col}} - \frac{1}{a_1}\right)}
       = \sqrt{\frac{\mu(1-e_1)}{a_{\rm col}}},
       \qquad \text{(particle~1 at apoapsis)} \\
  v_2 &= \sqrt{\mu\left(\frac{2}{a_{\rm col}} - \frac{1}{a_2}\right)}
       = \sqrt{\frac{\mu(1+e_2)}{a_{\rm col}}}.
       \qquad \text{(particle~2 at periapsis)}
\end{align}
Both velocities are directed tangentially in the prograde direction.
Since $e_1, e_2 \ge 0$, we always have $v_1 \le v_\mathrm{circ} \le v_2$,
where $v_\mathrm{circ} = \sqrt{\mu/a_{\rm col}}$ is the local circular velocity.
Hence particle~2, being faster, catches up with particle~1 from behind.
To determine the post-collision orbital elements,
we next specify the collision model.

We adopt the standard normal coefficient of restitution. Let $f_{\rm mass} \equiv m_1/m_2$ be the mass ratio of the two particles. For simplicity, we adopt equal masses ($f_{\rm mass}  = 1$) throughout. The center-of-mass velocity is $v_\mathrm{com} = \frac{f_{\rm mass}  \, v_1 + v_2}{f_{\rm mass}  + 1}$ and the relative velocity before the collision is $\Delta v = v_2 - v_1 > 0$.

In the center-of-mass frame, the particles approach each other with speeds
$\Delta v / (f_{\rm mass} +1)$ and $f_{\rm mass} \, \Delta v / (f_{\rm mass} +1)$, respectively.
After the collision, the relative velocity reverses direction and is reduced
by the normal coefficient of restitution $\varepsilon_n \in [0, 1]$ as $\Delta v' = \varepsilon_n \, \Delta v$.
Here $\varepsilon_n = 1$ corresponds to a perfectly elastic collision and $\varepsilon_n = 0$ to a perfectly inelastic collision or merge at $v_\mathrm{com}$. The fraction of relative kinetic energy dissipated is $1 - \varepsilon_n^2$.

The post-collision velocities in the planet-centered frame are:
\begin{align}
  v_1' &= v_\mathrm{com} + \frac{\varepsilon_n \, \Delta v}{f_{\rm mass} +1}, \\
  v_2' &= v_\mathrm{com} - \frac{f_{\rm mass}  \, \varepsilon_n \, \Delta v}{f_{\rm mass} +1}.
\end{align}
One can verify that momentum is conserved: $m_1 v_1' + m_2 v_2' = m_1 v_1 + m_2 v_2$.
Particle~1 (the slower particle that was caught from behind) is
accelerated ($v_1' > v_1$),
while particle~2 (the faster particle that caught up) is
decelerated ($v_2' < v_2$).

After the collision, each particle has a purely tangential velocity $v'$ at distance $a_{\rm col}$.
The new semi-major axis is obtained from the vis-viva equation:
\begin{equation}
  a' = \frac{1}{\dfrac{2}{a_{\rm col}} - \dfrac{v'^2}{\mu}}.
\end{equation}

An interesting feature emerges from this formulation. Whether $a_{\rm col}$ becomes the apoapsis or periapsis of the post-collision orbit depends on its comparison with the local circular velocity $v_\mathrm{circ} = \sqrt{\mu/a_{\rm col}}$. If $v' < v_\mathrm{circ}$, $a_{\rm col}$ corresponds to the apoapsis of the new orbit, whereas if $v' \ge v_\mathrm{circ}$, it corresponds to the periapsis. Accordingly, the post-collision eccentricity $e'$ is given by
\begin{equation}
e' =
\begin{cases}
\displaystyle \frac{a_{\rm col}}{a'} - 1 & \text{if $v' < v_\mathrm{circ}$,} \\[6pt]
\displaystyle 1 - \frac{a_{\rm col}}{a'} & \text{if $v' \ge v_\mathrm{circ}$.}
\end{cases}
\end{equation}

An important feature of the post-collision dynamics is whether a particle
crosses from the apoapsis constraint curve to the periapsis constraint curve
in the $a$--$e$ diagram. This crossing occurs when the post-collision velocity exceeds $v_\mathrm{circ}$. For particle~1 (initially at apoapsis with $v_1 < v_\mathrm{circ}$),
crossing requires $v_1' > v_\mathrm{circ}$, i.e.,
\begin{equation}
  v_\mathrm{com} + \frac{\varepsilon_n \, \Delta v}{f_{\rm mass} +1} > v_\mathrm{circ}.
\end{equation}
For equal masses ($f_{\rm mass}=1$), this gives a critical coefficient of restitution:
\begin{equation}
  \varepsilon_n > \varepsilon_n^\mathrm{crit}
    = \frac{v_\mathrm{circ} - v_\mathrm{com}}{\Delta v / 2}
    = \frac{2(v_\mathrm{circ} - v_\mathrm{com})}{v_2 - v_1}.
\end{equation}
When $v_\mathrm{com} < v_\mathrm{circ}$ (which occurs when the eccentricities are
sufficiently large), collisions with $\varepsilon_n < \varepsilon_n^\mathrm{crit}$
leave particle~1 on the apoapsis constraint curve,
while those with $\varepsilon_n > \varepsilon_n^\mathrm{crit}$ push it across to
the periapsis curve.
Figure~\ref{Fig_evolution_analytical} illustrates this behavior for several combinations
of initial eccentricities and $\varepsilon_n$.

Figure~\ref{Fig_evolution_analytical} shows that, regardless of the initial combinations of orbital elements ($a$ and $e$) and the value of $\varepsilon_n$, the particles exchange positions and the post-impact $a$ and $e$ follow one of the original V-shaped constraint curves while gradually decreasing $e$.
This result has a direct implication for the potential ring formation discussed in \citet{Teo23,Cuk26}. The collision outcome between particles belonging to different arms does not follow the evolutionary path assumed in the equivalent circular orbit approach. Instead, the particles remain confined to the original V-shaped structure. Regardless of the value of the normal coefficient of restitution $\varepsilon_n$, the debris orbits evolve toward the original impact radius as circularization proceeds.

In other words, even if debris particles initially possess $a$-$e$ combinations that allow them to pass within the Roche limit, their orbital elements rapidly evolve during successive collisional interactions. As a result, they do not remain inside the Roche limit long enough to contribute to ring formation. This demonstrates that the estimates of \citet{Teo23,Cuk26}, which rely solely on the equivalent circular orbit argument, neglect the essential long-term collisional evolution of the debris cloud. When this evolution is properly taken into account, the debris cloud does not form rings but instead converges toward orbits associated with the original impact location during circularization. 

We also directly compare the above analytical arguments with the direct two-body collisional simulations including fragmentation (Appendix \ref{appendix_twobody}; see also Appendix \ref{appendix_one_arm}), and confirmed the validity of the above arguments. Furthermore, in the next section, we demonstrate that this conclusion is not only supported by our analytical framework but is also confirmed by direct $N$-body simulations, including fragmentation.\\

\section{Numerical Approach}
\label{sec_numerical}

\subsection{Numerical Settings}

To test the analytical arguments developed in Section~\ref{sec_analytical}, we perform direct $N$-body simulations of the long-term evolution of the post-impact debris clouds, including collisional fragmentation \citep{Iwa20,Ish21}. Mutual gravitational interactions among particles are calculated using a particle--particle integration scheme \citep{Osh11}. Saturn is modeled as an axisymmetric central potential including the $J_2$ term, and the perturbation from Titan is also included. Physical collisions are detected when the separation between two particles becomes smaller than the sum of their radii.

For each collision, the outcome is determined by comparing the specific impact energy with the catastrophic disruption threshold appropriate for gravity-dominated bodies \citep{Cha13}. Depending on the impact conditions, collisions are classified as merging, hit-and-run, or fragmentation events. In the fragmentation regime, the mass of the largest remnant is estimated using standard collisional scaling relations \citep{Ste09,Lei12}, and the remaining mass is redistributed into a limited number of fragments (e.g., 10), with a minimum fragment radius of $\sim 24$\,km, in order to keep the calculations computationally tractable. Particles are removed from the system if they collide with Saturn or migrate beyond a prescribed outer boundary.

We performed three sets of simulations with different initial conditions, summarized in Table~\ref{tab:init_cond}. The initial total mass of the debris particles is set to be twice the mass of Rhea, $\sim 4 \times 10^{21}\,\mathrm{kg}$. The debris particles are initially modeled as equal-sized bodies with a radius of $200\,\mathrm{km}$ and are distributed uniformly over the adopted initial orbital-element distribution in the $a$--$e$ plane to represent the impact-generated debris population (Fig.~\ref{fig_schematic}).

All debris particles are generated at the same collision location and therefore initially share either a common periapse or a common apoapse, namely $q=a_{\rm col}$ or $Q=a_{\rm col}$, corresponding to the two arms of the V-shaped distribution discussed in Section~\ref{sec_analytical} (see Fig.~\ref{fig_schematic}). As an intentionally favorable initial condition for ring formation, the debris orbits are initialized such that their periapses can extend down to nearly Saturn's surface. In this sense, our setup is deliberately biased toward producing debris that could potentially contribute to Saturn's rings (see Fig.~\ref{fig_schematic}). Even under such favorable conditions, however, our simulations presented below show that a massive ring does not form.

We note that this initial orbital distribution is intentionally idealized. In a realistic giant impact, the released debris should have a finite radial velocity dispersion, and the corresponding distribution in the $a-e$ plane would therefore have a finite extension inside the idealized V-shaped constraint curves rather than lying exactly on them. In the present study, we do not impose such an additional dispersion artificially, because our goal is to compare the numerical calculations directly with the analytical model developed in Section \ref{sec_analytical}, and because a physically realistic dispersion would depend on the specific SPH impact outcome. We note, however, that in \cite{Hyo17d}, where SPH-generated debris was used as the initial condition for N-body simulations, such a finite dispersion was naturally included, yet the debris still reaccreted near the impact radius rather than forming a massive ring.

Because the inclination distribution of the debris is expected to depend sensitively on the original collision geometry between Saturnian proto-large moons, we explore several representative cases with different choices of the characteristic release distance $a_{\rm col}$ and a Gaussian inclination distribution characterized by a center $i$ and width $\sigma_i$ (Table~\ref{tab:init_cond}). These cases are intended to mimic plausible variations arising from different impact geometries of Saturnian proto-large moons \citep{Teo23,Cuk26}. A smaller inclination dispersion produces a geometrically thinner debris cloud, increases the collision rate, and accelerates the subsequent evolution.

We stop our simulations after $\sim 1$ or $\sim 10$ yr, depending on the case. By that time, the number of particles in the system has increased substantially because of collisional fragmentation, making further integration computationally prohibitive. Nevertheless, this timespan is sufficient to capture the fundamental physics of the V-shaped debris evolution and to demonstrate that the adopted initial conditions do not lead to the formation of a massive Saturnian ring (see Figs.~\ref{fig_nbody_as_evo} and \ref{fig_nbody_mass_evo}).

\begin{table}
\centering
\caption{Initial conditions adopted in the direct $N$-body simulations. The inclination distribution is generated from a Gaussian distribution and its absolute value is taken, so that $i \ge 0$. Here $R_{\rm Sat}$ is the radius of Saturn. The representative values are based on and motivated by previous studies \cite{Teo23,Cuk26}.}
\label{tab:init_cond}
\begin{tabular}{cccc}
\hline
\hline
Case & $a_{\rm col}$ & Inclination center & Inclination width \\
\hline
1 & $6R_{\rm Sat}$ & $5^\circ$ & $\sigma_i = 3^\circ$ \\
2 & $7R_{\rm Sat}$ & $5^\circ$ & $\sigma_i = 3^\circ$ \\
3 & $6R_{\rm Sat}$ & $0^\circ$ & $\sigma_i = 15^\circ$ \\
\hline
\end{tabular}
\end{table}

\subsection{Numerical Results}

\begin{figure*}[htbp]
	\begin{center}
	\includegraphics[width=0.85\textwidth]{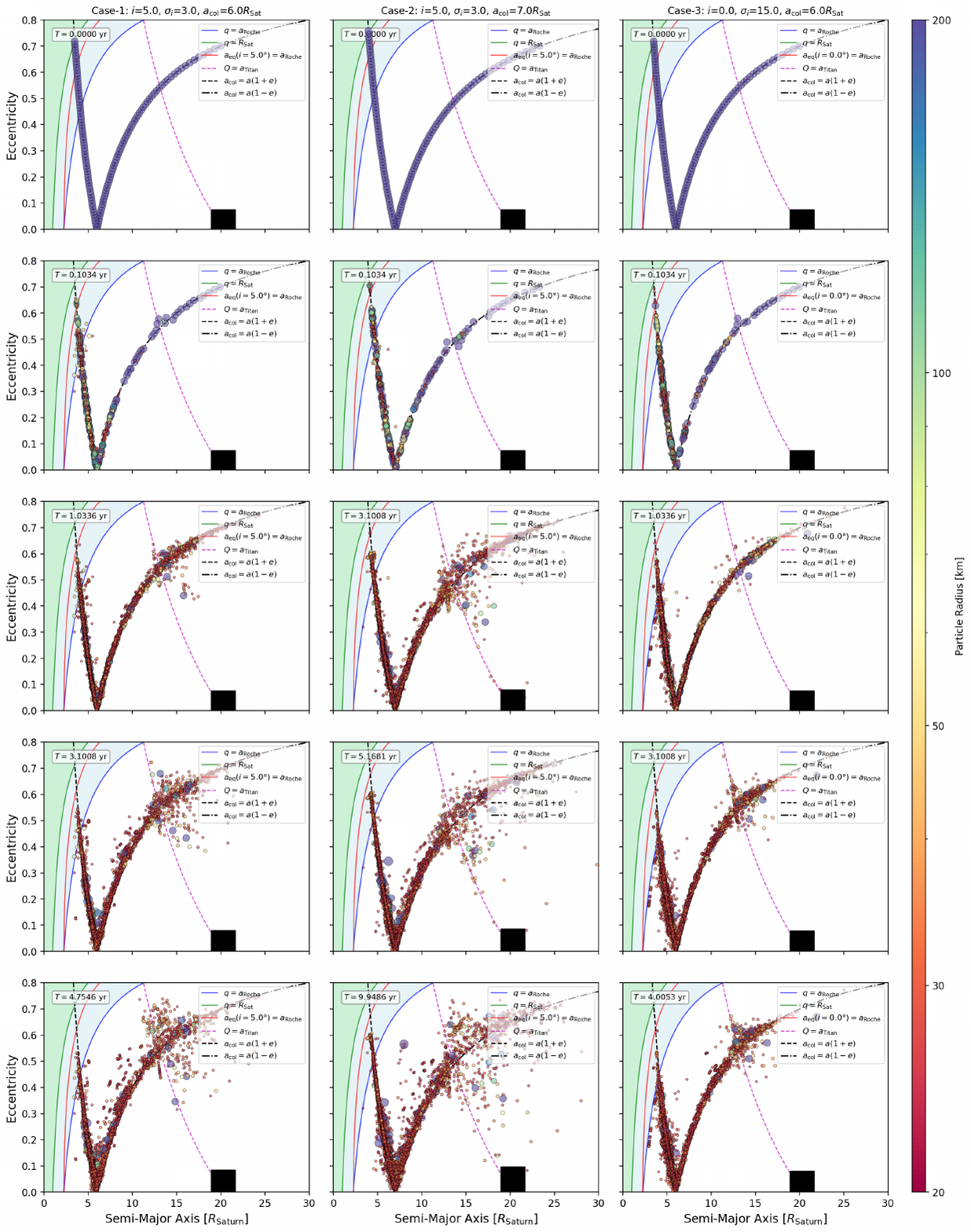}
	\caption{
Evolution of the debris distribution in the $a$--$e$ plane obtained from the direct $N$-body simulations for the three representative initial conditions summarized in Table~\ref{tab:init_cond}. Different columns correspond to different initial conditions, and different rows show snapshots at different times. The black curves indicate the original V-shaped constraint curves, $a_{\rm col}=a(1-e)$ and $a_{\rm col}=a(1+e)$, corresponding to debris particles that initially share a common periapsis or apoapsis at the collision radius $a_{\rm col}$. The blue and red curves mark the conditions $q=a_{\rm Roche}$ and $a_{\rm eq}=a_{\rm Roche}$, respectively, where $q=a(1-e)$ is the periapsis distance and $a_{\rm eq}$ is the equivalent circular radius. Green curves indicate $q=R_{\rm Sat}$. The color and size of each point indicates the fragment size. Titan is plotted as a black square.
}
	\label{fig_nbody_as_evo}
	\end{center}
\end{figure*}

Figure~\ref{fig_nbody_as_evo} shows the evolution of the debris distribution in the $a$--$e$ plane from our direct $N$-body simulations for several representative cases (different columns correspond to different initial conditions). At $t=0$, immediately after the impact, the debris is distributed over a wide V-shaped region in orbital-element space, and some debris particles have periapses inside the Roche limit ($q < a_{\rm Roche}$). In addition, some debris particles initially satisfy $a_{\rm eq} < a_{\rm Roche}$. These are the debris particles that would be counted as potential ring material under the equivalent-circular-orbit interpretation adopted by \citet{Teo23,Cuk26}.

However, once the long-term collisional evolution is followed explicitly, the mass residing in these regions decreases rapidly. Soon after the start of the $N$-body integration, only a small fraction of the debris still satisfies either $q < a_{\rm Roche}$ or $a_{\rm eq} < a_{\rm Roche}$. This behavior arises because the debris does not evolve at fixed angular momentum toward isolated circular orbits. Instead, as shown by the analytical arguments in Section~\ref{sec_analytical}, mutual collisions among debris particles immediately modify their orbital elements, and the debris evolves approximately along the original V-shaped constraint curves toward the apex of the V-shape, i.e., toward the impact radius. 

Although the initial conditions are placed on the idealized V-shaped curves, the particle distribution develops a finite width as the system evolves through repeated collisions and fragmentation (Fig.~\ref{fig_nbody_as_evo}). In this sense, the simulations naturally produce debris located inside the idealized V-shape as well. Nevertheless, this broadening does not alter the qualitative outcome: the debris does not evolve toward a long-lived massive ring inside the Roche limit, but instead continues to undergo collisional damping and converges toward the apex of the original V-shaped structure.

Comparing the different cases (Table~\ref{tab:init_cond}), the left and middle panels correspond to cases with the same inclination distribution but different initial collision radii, $a_{\rm col}=6R_{\rm Sat}$ and $7R_{\rm Sat}$, respectively. A larger $a_{\rm col}$ results in a longer orbital period and hence a longer collisional evolution timescale. The left and right panels, on the other hand, correspond to cases with the same $a_{\rm col}$ but different inclination distributions. Although their initial orbital distributions differ, both cases show rapid collisional evolution accompanied by eccentricity damping of particles that initially satisfy either $q < a_{\rm Roche}$ or $a_{\rm eq} < a_{\rm Roche}$, and the mass in these regions is quickly depleted.

\begin{figure*}[htbp]
	\begin{center}
	\includegraphics[width=0.5\textwidth]{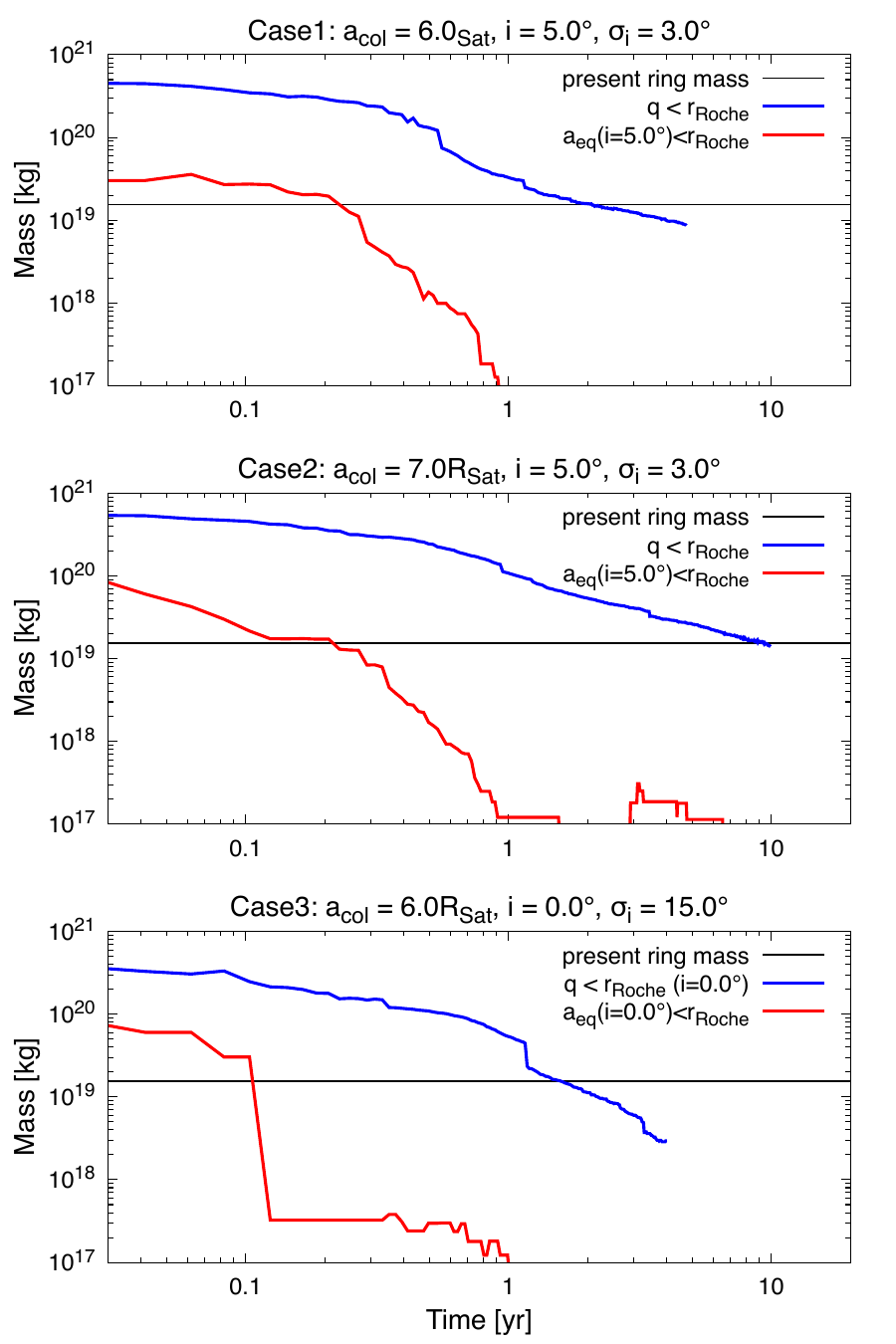}
	\caption{
Time evolution of the total mass of debris satisfying $q < a_{\rm Roche}$ (blue curves) and $a_{\rm eq} < a_{\rm Roche}$ (red curves) in the direct $N$-body simulations for the three representative initial conditions. The present mass of Saturn's rings is shown by black line. Here, $q=a(1-e)$ is the periapsis distance and $a_{\rm eq}$ is the equivalent circular radius. At $t=0$, the mass satisfying $a_{\rm eq} < a_{\rm Roche}$ is sufficiently large to account for the present-day mass of Saturn's rings, as assumed in the equivalent-circular-orbit interpretation. However, once the actual collisional evolution is followed, this mass rapidly decreases by orders of magnitude. The mass satisfying $q < a_{\rm Roche}$ also decreases with time, but less drastically. 
}
	\label{fig_nbody_mass_evo}
	\end{center}
\end{figure*}

This behavior is more clearly seen in Figure~\ref{fig_nbody_mass_evo}, which shows the corresponding mass evolution. The total mass of debris satisfying $q < a_{\rm Roche}$ (blue curves) decreases with time. The mass satisfying $a_{\rm eq} < a_{\rm Roche}$ (red curves) decreases even more drastically. This stronger decrease is expected because the condition $a_{\rm eq} < a_{\rm Roche}$ corresponds to a more restrictive region in the $a$--$e$ plane, located effectively closer to Saturn than the simple periapse condition $q < a_{\rm Roche}$ (see the corresponding curves in Fig.~\ref{fig_nbody_as_evo}). The blue curves therefore directly quantify how the equivalent-circular-orbit estimate evolves once the actual collisional dynamics is taken into account.

At $t=0$, the mass inferred from the condition $a_{\rm eq} < a_{\rm Roche}$ is large enough to match the present mass of Saturn's rings (approximately $1.54 \times 10^{19}\ {\rm kg}$; \cite{Ies19}), consistent with the interpretation of \citet{Teo23,Cuk26}. However, this estimate is valid only if one ignores the subsequent collisional evolution of the debris. Our direct $N$-body simulations show that, over longer timescales, the mass associated with $a_{\rm eq} < a_{\rm Roche}$ decreases by orders of magnitude. Consequently, once the actual dynamical evolution is followed, the surviving mass is far too small to account for the present-day mass of Saturn's rings.

Comparing the mass-evolution timescales among the different cases (Table~\ref{tab:init_cond}), a larger $a_{\rm col}$ (e.g., $a_{\rm col}=7R_{\rm Sat}$; middle panel) results in a longer orbital period and therefore tends to produce a longer collisional evolution timescale. However, the exact scaling is not determined by the orbital period alone, because the collision rate also depends on the spatial distribution of the debris and on the stochastic onset of fragmentation, which can strongly modify the number of particles and hence the subsequent collision probability. Nevertheless, for all parameter combinations considered in this study, the total mass satisfying $q < a_{\rm Roche}$ (blue curves) falls below the present mass of Saturn's rings within $\sim 10$~yr. More importantly, the total mass satisfying $a_{\rm eq} < a_{\rm Roche}$ (red curves), which has been used in previous studies as an estimate of the potential ring mass, falls below the present ring mass within only $\sim 0.1$~yr. Thus, although the precise evolution timescale depends on the initial conditions, the rapid depletion of the debris identified as potential ring material is a robust outcome.

These numerical results confirm the analytical conclusion: although some debris initially penetrates the Roche limit on eccentric orbits, the post-impact debris cloud does not evolve into a massive ring through simple circularization at constant angular momentum. Instead, the debris remains constrained by the original V-shaped structure and tends to converge near the original impact location.

We also find that, by the end of our simulations, fragments with relatively small eccentricities near the apex of the V-shape begin to reaccrete and form new moons. A qualitatively similar evolutionary pathway was reported in the previous $N$-body simulations of \citet{Hyo17d}, although their adopted initial V-shaped distribution was narrower than in the present study and their calculations did not include collisional fragmentation physics. Our results therefore suggest that, even when collisional fragmentation is taken into account, catastrophic collisions between Saturnian proto-large moons do not produce massive rings, but instead lead to the reaccretion of debris into a new generation of moons \citep[see also][]{Hyo17d}.

Much longer simulations that fully track the reaccretion of the debris are computationally very demanding, even with the use of supercomputer clusters. In the present study, our focus was on the fundamental dynamical evolution of V-shaped impact debris clouds and on whether such debris can form Saturn's rings. We therefore leave a full investigation of the final reaccretion process to future work.

\section{Conclusions and Future Perspectives}
\label{sec_summary}

In this study, we re-examined the long-term dynamical fate of debris clouds produced by catastrophic collisions between Saturnian proto-large moons. \citet{Teo23} and \citet{Cuk26} argued that a sufficient mass to form Saturn's rings could be supplied. However, this interpretation relies on the assumption that the post-impact debris circularized at fixed angular momentum, such that the resulting equivalent circular radius lay inside the Roche limit, while neglecting the actual collisional evolution of the debris clouds. The post-impact debris is not distributed along a single branch in the $a$--$e$ plane, but instead forms a characteristic V-shaped structure composed of two distinct arms corresponding to debris particles that share a common impact radius \citep{Hyo17d,Teo23}. Our analytical arguments and direct $N$-body simulations demonstrate that, because particles on the two arms possess significantly different angular momenta, inter-arm collisions dominate the evolution and fundamentally alter the orbital elements of the debris. As a result, the equivalent circular orbit approach cannot be used to assess the long-term fate of the collision debris.

Our numerical simulations confirm this analytical picture. Although some debris particles initially pass within the Roche limit on eccentric orbits, their subsequent collisional evolution does not lead to simple inward circularization into a massive ring. Instead, the debris remains approximately confined to the original V-shaped constraint curves and evolves toward the apex of the V-shape, i.e., toward the original collision location. Consequently, the mass that would be inferred to lie inside the Roche limit from the equivalent circular radius decreases rapidly with time once the actual dynamical evolution is followed. 

Accordingly, the ring-formation interpretation of \citet{Teo23} and \citet{Cuk26} should be reconsidered in light of the coupled collisional and dynamical evolution demonstrated here\footnote{We note, however, that our study is not intended to assess other aspects of the scenarios proposed by \citet{Teo23} and \citet{Cuk26}, such as the recent catastrophic impact scenario itself and the ancient resonant configuration. Those aspects may still remain valid.}. Under the conditions considered here, such collisions do not produce a massive ring. Rather, the debris dynamically evolves toward reaccretion near the original impact radius. Thus, the formation of a new generation of moons, as proposed by \citet{Teo23} and \citet{Cuk26}, remains a valid outcome and is in fact supported by our simulations. This conclusion is also qualitatively consistent with the earlier $N$-body study of \citet{Hyo17d}, although that study did not include collisional fragmentation physics. We note, however, that the present study does not explicitly test the additional pathway discussed by \citet{Teo23}, in which debris on the inner arm of the V-shape may be scattered by inner moons onto lower-angular-momentum orbits. Although inner moons such as Tethys, Enceladus, and Mimas are not included here, Titan is already included in our N-body simulations and the debris identified as potential ring material is still rapidly depleted. Our results therefore suggest that, to substantially alter the outcome, any such additional scattering pathway would need to operate on similarly short timescales and have a sufficiently strong effect on the debris evolution. 

A full treatment of the final reaccretion phase, in which the fragments are followed until they merge into one or more large moons, requires substantially longer integrations and is computationally very demanding, even with the use of supercomputer clusters. In the present work, our primary objective was to clarify the fundamental dynamical evolution of V-shaped impact debris clouds and to determine whether such debris can form Saturn's rings. We have shown that, under the conditions considered here, it does not lead to the formation of a massive Saturnian ring. Following the complete reaccretion process in detail, including the final growth history and architecture of the reaccreted moons, is therefore left for future work.

More broadly, our study provides a revised dynamical framework for understanding the evolution of collision-generated debris clouds in planetary systems. The key point is that, when the debris cloud is generated with a broad V-shaped distribution in orbital-element $a$--$e$ space, its subsequent evolution is controlled by inter-arm collisions and cannot be approximated by independent circularization at fixed angular momentum. Moreover, even in the case of a one-armed debris distribution, independent circularization at fixed angular momentum may not be an appropriate approximation, because collisional redistribution can couple particles across a broad range of eccentricities and semimajor axes (see Appendix~\ref{appendix_twobody} and Appendix~\ref{appendix_one_arm} for further details).

Therefore, the dynamical framework developed in this study for V-shaped impact debris clouds should be applicable to a broader range of collisional phenomena in planetary systems. Similar V-shaped debris distributions have been identified in giant-impact events, such as the impact thought to have formed the Moon around the Earth \citep{Bot15} and the giant impact proposed for the formation of the Martian moons \citep{Hyo18}. The dynamical understanding obtained here may therefore provide a useful framework for interpreting the fate of impact-generated debris clouds in a wide variety of planetary collision environments (see also Appendix~\ref{appendix_one_arm}).

\begin{acknowledgments}
R.H. acknowledges the financial support of JSPS Grants-in-Aid (26K00756, 23KK0253, 22K14091, 21H04512, 21H04514, 20KK0080). N.T. acknowledges the financial support of JSPS KAKENHI Grant Number 25KJ1252. We thank Shigeru Ida and Sebastien Charnoz for discussion.
\end{acknowledgments}

\newpage
\appendix
\setcounter{figure}{0}%
\renewcommand{\thefigure}{A\arabic{figure}}%
\section{$N$-body Simulation of a Two-Particle Collision for Comparison with the Analytical Model}
\label{appendix_twobody}

To directly compare with the analytical framework developed in Section~\ref{sec_analytical}, 
we perform a simplified $N$-body experiment considering only two initial particles. 
The particles are initially placed on the two arms of the V-shaped distribution, 
such that their apoapsis and periapsis coincide at the common collision radius $a_{\rm col}$. 
This setup allows a one-to-one comparison with the analytical two-body treatment.

In order to introduce additional physical complexity beyond the purely analytical model, 
we include gravity among debris particles and collisional fragmentation in the numerical simulations. 
When the specific impact energy exceeds the catastrophic disruption threshold, 
the colliding bodies are partially disrupted and fragments are generated 
following the adapted scaling relations. The simulations are stopped slightly after the first catastrophic impact between particles initially located on different arms. 

\begin{figure*}[htbp]
	\begin{center}
	\includegraphics[width=0.8\textwidth]{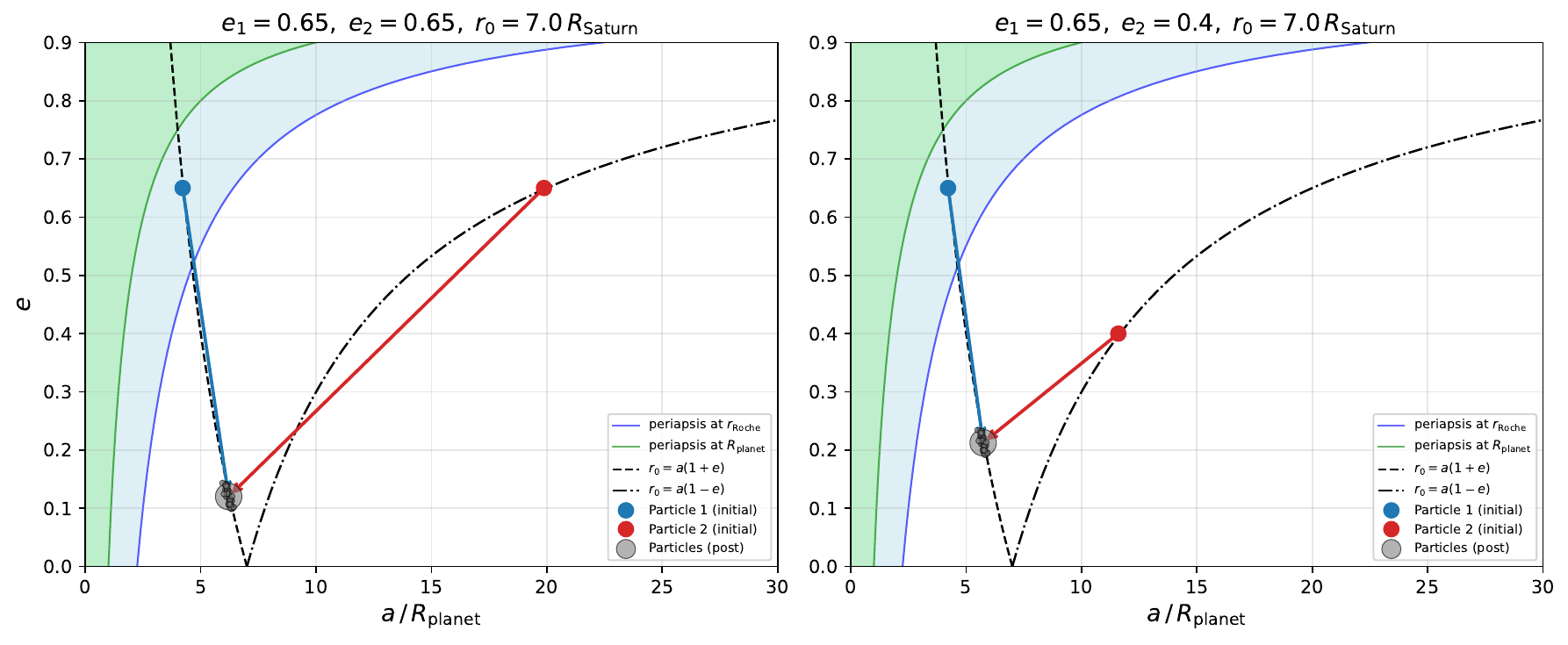}
	\caption{Results of direct $N$-body simulations including fragmentation. 
Only two initial particles are considered, placed on the two arms of the 
V-shaped distribution so that their apoapsis and periapsis coincide at $a_{\rm col}$. 
The simulations are stopped slightly after the first catastrophic impact 
between particles initially located on different arms (i.e., blue and red particles). 
Following the catastrophic collision, the post-impact orbital elements of fragments $(a,e)$ 
shift to new positions that remain confined to the original constraint curves 
defined by $a_{\rm col} = a(1 \pm e)$. This behavior is fully consistent with the analytical framework (Section~\ref{sec_analytical}; Figure~\ref{Fig_evolution_analytical}). 
Although secondary collisions among fragments sometimes occur before the simulation is stopped, these fragment-fragment interactions involve significantly smaller relative 
velocities compared to the initial inter-arm catastrophic collision, 
and therefore do not substantially modify the $a$-$e$ evolution. 
The numerical results demonstrate that even when fragmentation and secondary 
impacts are included, the debris evolution remains governed by the same 
V-shaped dynamical constraint predicted analytically.}
	\label{figA1_twobody_numerical}
	\end{center}
\end{figure*}

As shown in Fig.~\ref{figA1_twobody_numerical}, even when fragmentation and subsequent fragment-fragment collisions are included, the post-impact orbital elements remain confined to the original V-shaped constraint curves defined by $a_{\rm col} = a(1 \pm e)$. 
Although some secondary impacts among fragments occur, their relative velocities are 
significantly smaller than that of the initial inter-arm catastrophic collision, 
and therefore rarely modify the $a$-$e$ evolution. 

Here, after the impacts, we observed the energy dissipation is $ |(E_{\rm after}-E_{\rm initial})/E_{\rm initial}| \sim 0.9$, where $E_{\rm after}$ and $E_{\rm initial}$ are the total energy of the system after and before the collisions, respectively. This indicates the corresponding effective $\varepsilon_{\rm n,eff}$  is $\sim 0.3$ as the energy dissipation is $1-\varepsilon_{\rm n,eff}^2$. This effective value, $\varepsilon_{n,{\rm eff}} \sim 0.3$, is qualitatively consistent with the strongly dissipative cases in Fig.~\ref{Fig_evolution_analytical}, such as the $\varepsilon_n \sim 0.1$ case. The methodological difference from the analytical arguments, for example, inclusion of fragmentation leads to the clumping of its fragments at a very similar $a$-$e$ position among fragments. Overall, the numerical results confirm that the debris evolution is governed by the same dynamical constraint derived analytically, demonstrating the robustness of the analytical arguments even when collisional fragmentation is taken into account.

\section{$N$-body Simulation of One Arm}
\label{appendix_one_arm}

To isolate the role of same-arm collisions and to better understand the fundamental dynamics of the V-shaped debris distribution, we perform six additional idealized $N$-body experiments in which particles are initially placed on only one branch of the V-shape. The first three cases, presented in Section~\ref{appendix_one_arm_inner}, correspond to the inner-arm configuration in which all particles share a common apoapsis, $Q=a_{\rm col}$ (Fig.~\ref{figA2_nbody_numerical}). The next three cases, presented in Section~\ref{appendix_one_arm_outer}, correspond to the outer-arm configuration in which all particles share a common periapsis, $q=a_{\rm col}$ (Fig.~\ref{figA3_nbody_numerical}). The inner-arm and outer-arm subsections (Sec.~\ref{appendix_one_arm_inner} and Sec.~\ref{appendix_one_arm_outer}) each consider three idealized initial distributions, labeled setup-1 through setup-3 and setup-4 through setup-6, respectively: a single localized cluster, a homogeneous one-arm distribution, and two localized clusters placed at different positions along the same branch. The $J_2$ term is not included here.

\subsection{Case of only the inner arm}
\label{appendix_one_arm_inner}

In setup-1, the debris particles are initially concentrated within a narrow region of $a$--$e$ space and therefore have similar specific angular momenta. In this case, successive collisional damping, including fragmentation, is expected to dissipate orbital energy while conserving the total angular momentum of the debris cloud. Because the particles initially share similar angular momentum, the longer-term evolution in the $a$--$e$ plane is expected to approach a constant-angular-momentum-like locus, for which the equivalent circular radius provides a useful reference. We therefore plot the corresponding constant $a_{\rm eq}$ contour in Fig.~\ref{figA2_nbody_numerical}. Although the simulations begin to show this tendency, with the debris distribution starting to deviate from the initial negatively sloped inner-arm branch toward a more positively sloped constant-angular-momentum-like locus, a direct verification of the much longer-term evolution would require substantially longer integrations and is computationally prohibitive in the present framework.

In setup-2, the debris particles are initially distributed over a wide range of semi-major axes along a single branch. Because the synodic period becomes shorter for particle pairs with larger $|\Delta a|$, collisions first occur preferentially between debris particles that are widely separated along the branch. These collisions produce fragments whose orbital elements are redistributed toward intermediate $a$--$e$ values, and the particles therefore first concentrate around the mean location of the initial distribution (approximately $e \sim 0.4$ in Fig.~\ref{figA2_nbody_numerical}). Once the fragments become clustered in a narrower range of $a$, the subsequent collisional evolution slows down because the synodic period becomes longer. At later times, the clustered fragments would begin to spread preferentially along a locus approximately close to constant angular momentum in the $a$--$e$ plane, as we observe in setup-1.

In setup-3, two clusters are initially placed at different $a$--$e$ locations along the same branch. In this case, the evolution is intermediate between setup-1 and setup-2. First, collisions between particles belonging to different clusters occur on a relatively short timescale and generate a new concentration near the mean $a$--$e$ value of the two initial clusters (approximately $e \sim 0.4$ in Fig.~\ref{figA2_nbody_numerical}). Subsequently, collisions within and among these concentrations further redistribute the fragments, and the particle distribution becomes progressively more continuous along the $Q=a_{\rm col}$ branch. At late times, the resulting distribution approaches a more extended one-arm configuration, qualitatively similar to that in setup-2, although the particle size distribution has already been modified by collisional fragmentation. Therefore, over longer timescales, setup-3 is expected to follow an evolutionary path similar to that of setup-2.

Overall, these additional experiments provide further physical insight into the fundamental dynamical evolution of impact debris. In particular, the direct $N$-body results are fully consistent with the analytical arguments presented in the main text: the early collisional evolution is primarily controlled by the synodic timescale, such that particle pairs with larger $|\Delta a|$ collide first, whereas later evolution slows down as the debris becomes concentrated in a narrower range of semi-major axis. 
 
 \subsection{Case of only the outer arm}
 \label{appendix_one_arm_outer}
 
\begin{figure*}[htbp]
	\begin{center}
	\includegraphics[width=0.85\textwidth]{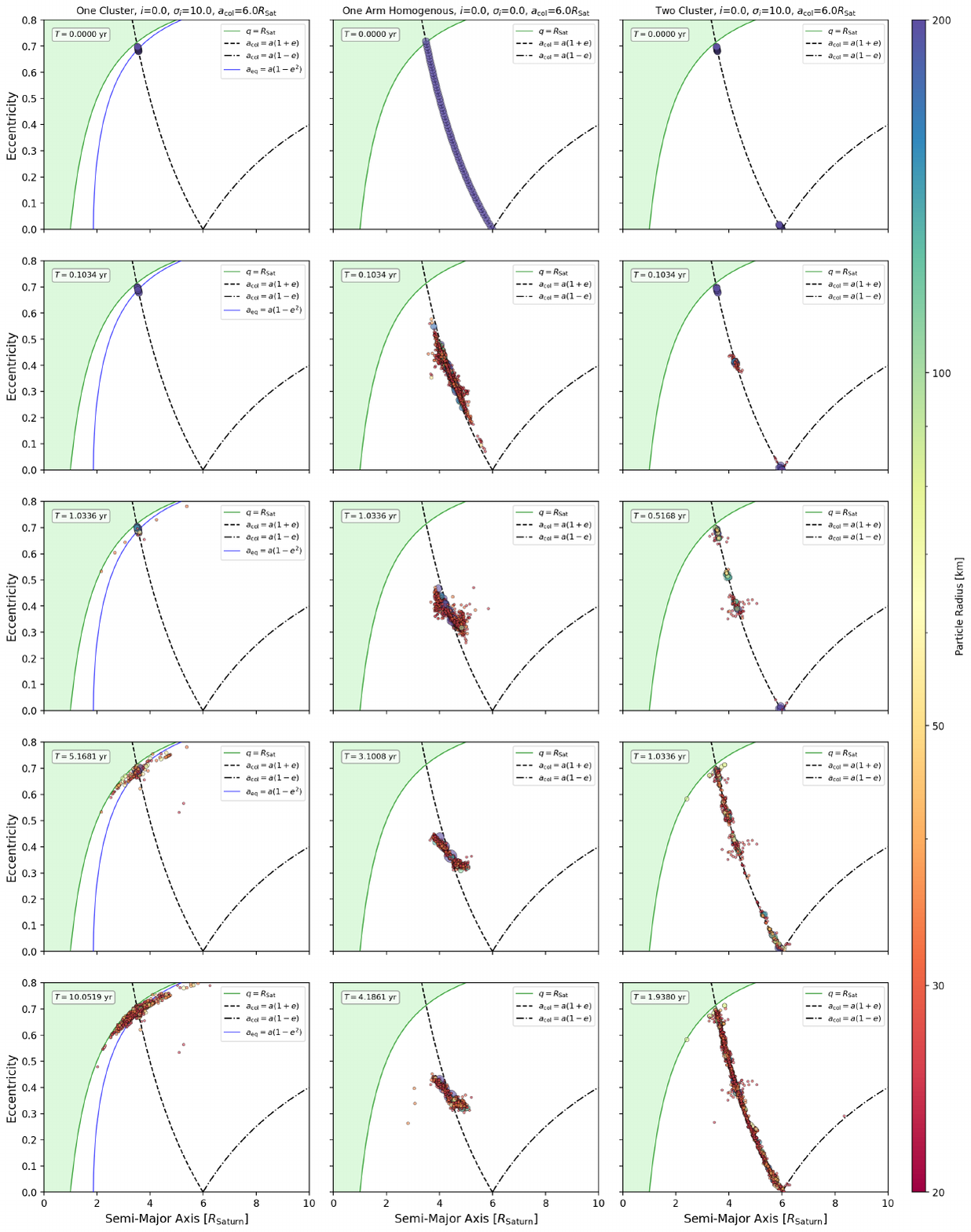}
	\caption{Results of direct $N$-body simulations including fragmentation for three idealized one-arm initial conditions. In all cases, particles are initially placed on a single branch of the V-shaped distribution and satisfy a common apoapsis, $Q=a_{\rm col}$ (black dashed lines). Green curves indicate $q=R_{\rm Sat}$. Left: setup-1, in which debris particles are initially concentrated in a single localized cluster on the same branch. The blue curve indicates the constant $a_{\rm eq}$, or equivalently constant-angular-momentum contour passing through the mean $(a,e)$ point of the initial cluster. The intersection of this curve with the $e=0$ axis
gives the corresponding equivalent circular radius $a_{\rm eq}=a(1-e^2)$ in the coplanar limit. Middle: setup-2, in which debris particles are uniformly distributed along the $Q=a_{\rm col}$ branch. Right: setup-3, in which two localized clusters are initially placed at two distinct locations along the $Q=a_{\rm col}$ branch.}
	\label{figA2_nbody_numerical}
	\end{center}
\end{figure*}

\begin{figure*}[htbp]
	\begin{center}
	\includegraphics[width=0.85\textwidth]{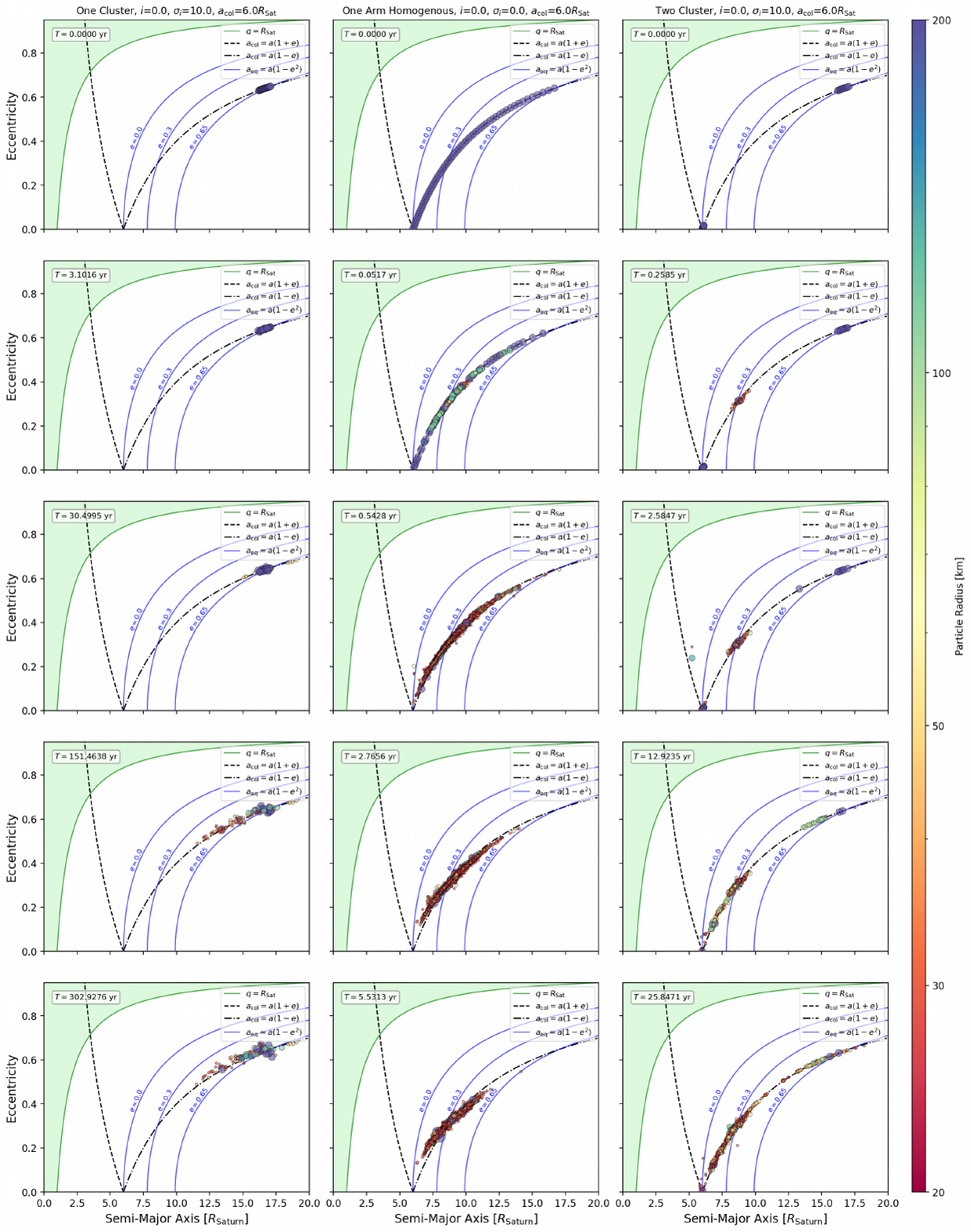}
	\caption{The same as Fig.~\ref{figA2_nbody_numerical}, but the case of the outer one-arm initial conditions. Left: setup-4, in which debris particles are initially concentrated in a single localized cluster on the same branch. Middle: setup-5, in which debris particles are uniformly distributed along the $q=a_{\rm col}$ branch. Right: setup-6, in which two localized clusters are initially placed at two distinct locations along the $q=a_{\rm col}$ branch. The blue curves indicate constant $a_{\rm eq}$, or equivalently constant-angular-momentum contours passing through the reference points $(a,e)$ on the $a_{\rm col}=a(1-e)$ branch with $e=0.0$, 0.3, and 0.65, respectively. The intersections of these curves with the $e=0$ axis give the corresponding equivalent circular radii $a_{\rm eq}=a(1-e^2)$ in the coplanar limit.}
	\label{figA3_nbody_numerical}
	\end{center}
\end{figure*}

Figure~\ref{figA3_nbody_numerical} shows the same set of one-arm calculations as Figure~\ref{figA2_nbody_numerical}, but for the outer branch of the V-shaped distribution, defined by a common periapsis distance, $q=a(1-e)=a_{\rm col}$. Such a configuration can arise, for example, from giant impacts \citep[e.g.,][]{Can04,Hyo17a,Hyo17b}\footnote{The debris distribution produced by the Martian giant impact \citep{Hyo17a,Hyo17b} is not, in general, distributed along a constant-angular-momentum or constant $a_{\rm eq}$ locus. Rather, it is closer to a distribution with a nearly common periapsis, i.e., approximately along a $q=a(1-e)$ branch in the $a$--$e$ plane. The debris distribution in the Moon-forming giant-impact model \citep{Can04} appears to show a similar tendency. Thus, applying the equivalent-circular-radius argument to such debris distributions requires caution, although a detailed reassessment is beyond the scope of this work. By contrast, tidal disruption tends to produce debris distributed nearly along a constant-angular-momentum line in $a$--$e$ space \citep[][]{Hyo17c,Tor26}.}. As in Figure~\ref{figA2_nbody_numerical}, we consider three idealized initial conditions: setup-4 consists of a single localized cluster, setup-5 of a homogeneous one-arm distribution, and setup-6 of two localized clusters. For reference, the blue curves show constant $a_{\rm eq}$, or equivalently constant-angular-momentum contours passing through the reference points on the $q=a_{\rm col}$ branch with $e=0.0$, $0.3$, and $0.65$, respectively, where $a_{\rm eq}=a(1-e^2)$ in the coplanar limit. The fact that these blue curves intersect the outer arm at different locations indicates that the debris particles distributed along the arm have different values of $a_{\rm eq}$.

In all three setups, the debris evolution remains primarily confined to the original $q=a_{\rm col}$ branch, rather than following the blue constant $a_{\rm eq}$ reference contours (Fig.~\ref{figA3_nbody_numerical}). This result is physically important. The fact that the particle distribution does not evolve along the constant $a_{\rm eq}$ contours does not imply any violation of angular-momentum conservation for the system as a whole. On the contrary, the total angular momentum of the debris cloud is well conserved in our numerical simulations: it is conserved to within $\sim 0.1$\,\% in most cases, with a maximum error of $\sim 1$\,\%. The residual error is attributable to numerical errors associated with the collision treatment and orbital integration. Importantly, collisions and collisional fragmentation continuously redistribute angular momentum among individual particles and fragments. Therefore, the evolution of each particle cannot, in general, be described as independent circularization while retaining its initial angular momentum. Instead, because collisions repeatedly occur under the nearly shared-pericenter geometry, the post-collision fragments are preferentially redistributed onto orbits that still approximately satisfy the same geometric constraint,
\begin{equation}
	q \simeq a_{\rm col}.
\end{equation}
Accordingly, the early-to-intermediate evolution of the outer one-arm configuration is governed more strongly by the collision geometry and the collisional redistribution of angular momentum among particles than by independent evolution toward the corresponding equivalent circular radius. As the evolution proceeds, however, the shared-pericenter geometry gradually breaks down because of the scattering caused by mutual gravity and collisional fragmentation.

The single-cluster case (setup-4; left panels in Fig.~\ref{figA3_nbody_numerical}) is particularly instructive. Because the particles initially occupy a narrow region in orbital-element space and therefore have similar specific angular momenta, one might expect the debris to evolve toward the equivalent circular radius associated with the cluster. However, the direct $N$-body simulations show that, within the simulated timespan, the particle distribution remains much more closely tied to the common-periapsis branch than to the constant $a_{\rm eq}$ reference contour. This indicates that, even in this idealized case, the shared collision geometry provides a stronger dynamical constraint during the early collisional stage. Although conservation of the total angular momentum of the debris cloud is expected ultimately to lead to convergence toward a constant-angular-momentum-like locus, demonstrating such behavior would require substantially longer integrations and is beyond the scope of the present calculations. The evolution of the other cases (setup-5 and setup-6; middle and right panels in Fig.~\ref{figA3_nbody_numerical}) can be understood in a similar way to that in Sec.~\ref{appendix_one_arm_inner}: the synodic timescale determines which particles collide first, and the resulting collisions and fragmentation progressively redistribute angular momentum, driving the debris toward a more concentrated distribution in $a$--$e$ space while gradually breaking down the initial geometric alignment.

In summary, the debris cloud remains preferentially confined to the original branch over the simulated timespan. This further illustrates that, when the debris initially occupies a broad region in the $a$--$e$ plane, its subsequent evolution cannot, in general, be inferred by treating each debris particle as evolving independently toward its own equivalent circular radius based on its initial orbital elements. The equivalent circular radius may become a useful descriptor only in the limiting case where the debris particles share nearly the same angular momentum from the outset, whether because they are initially confined to a narrow region of orbital-element space or because they are distributed along a nearly constant-angular-momentum locus \citep[e.g.,][]{Hyo17c,Tor26}.

\newpage
\bibliography{reference}{}

@ARTICLE{Tor26,
       author = {{Torii}, Naoya and {Ida}, Shigeru and {Hyodo}, Ryuki},
        title = "{Ring formation around giant planets by tidal disruption of a single passing large Kuiper belt object II: The dynamical fate of tidal fragments}",
      journal = {\icarus},
         year = 2026,
        month = sep,
       volume = {455},
          eid = {117097},
        pages = {117097},
          doi = {10.1016/j.icarus.2026.117097},
archivePrefix = {arXiv},
       eprint = {2604.10042},
 primaryClass = {astro-ph.EP},
       adsurl = {https://ui.adsabs.harvard.edu/abs/2026Icar..45517097T}
}

@BOOK{Kau66,
       author = {{Kaula}, William M.},
        title = "{Theory of satellite geodesy. Applications of satellites to geodesy}",
         year = 1966,
       adsurl = {https://ui.adsabs.harvard.edu/abs/1966tsga.book.....K}
}

@ARTICLE{Wis22,
       author = {{Wisdom}, Jack and {Dbouk}, Rola and {Militzer}, Burkhard and {Hubbard}, William B. and {Nimmo}, Francis and {Downey}, Brynna G. and {French}, Richard G.},
        title = "{Loss of a satellite could explain Saturn{\textquoteright}s obliquity and young rings}",
      journal = {Science},
         year = 2022,
        month = sep,
       volume = {377},
       number = {6612},
        pages = {1285-1289},
          doi = {10.1126/science.abn1234},
       adsurl = {https://ui.adsabs.harvard.edu/abs/2022Sci...377.1285W}
}

@ARTICLE{Can04,
       author = {{Canup}, Robin M.},
        title = "{Simulations of a late lunar-forming impact}",
      journal = {\icarus},
         year = 2004,
        month = apr,
       volume = {168},
       number = {2},
        pages = {433-456},
          doi = {10.1016/j.icarus.2003.09.028},
       adsurl = {https://ui.adsabs.harvard.edu/abs/2004Icar..168..433C}
}

@ARTICLE{Can10,
       author = {{Canup}, Robin M.},
        title = "{Origin of Saturn's rings and inner moons by mass removal from a lost Titan-sized satellite}",
      journal = {\nat},
         year = 2010,
        month = dec,
       volume = {468},
       number = {7326},
        pages = {943-946},
          doi = {10.1038/nature09661},
       adsurl = {https://ui.adsabs.harvard.edu/abs/2010Natur.468..943C}
}

@article{Iwa20,
  title={Implementation and performance of Barnes-hut n-body algorithm on extreme-scale heterogeneous many-core architectures},
  author={Iwasawa, Masaki and Namekata, Daisuke and Sakamoto, Ryo and Nakamura, Takashi and Kimura, Yasuyuki and Nitadori, Keigo and Wang, Long and Tsubouchi, Miyuki and Makino, Jun and Liu, Zhao and others},
  journal={The International Journal of High Performance Computing Applications},
  volume={34},
  number={6},
  pages={615--628},
  year={2020},
  publisher={Sage Publications Sage UK: London, England}
}

@article{Ish21,
  title={Particle--particle particle--tree code for planetary system formation with individual cut-off method: GPLUM},
  author={Ishigaki, Yota and Kominami, Junko and Makino, Junichiro and Fujimoto, Masaki and Iwasawa, Masaki},
  journal={Publications of the Astronomical Society of Japan},
  volume={73},
  number={3},
  pages={660--676},
  year={2021},
  publisher={Oxford University Press}
}

@ARTICLE{Osh11,
       author = {{Oshino}, Shoichi and {Funato}, Yoko and {Makino}, Junichiro},
        title = "{Particle-Particle Particle-Tree: A Direct-Tree Hybrid Scheme for Collisional N-Body Simulations}",
      journal = {\pasj},
         year = 2011,
        month = aug,
       volume = {63},
        pages = {881},
          doi = {10.1093/pasj/63.4.881},
archivePrefix = {arXiv},
       eprint = {1101.5504},
 primaryClass = {astro-ph.EP},
       adsurl = {https://ui.adsabs.harvard.edu/abs/2011PASJ...63..881O}
}

@ARTICLE{Cha13,
       author = {{Chambers}, J.~E.},
        title = "{Late-stage planetary accretion including hit-and-run collisions and fragmentation}",
      journal = {\icarus},
         year = 2013,
        month = may,
       volume = {224},
       number = {1},
        pages = {43-56},
          doi = {10.1016/j.icarus.2013.02.015},
       adsurl = {https://ui.adsabs.harvard.edu/abs/2013Icar..224...43C}
}

@ARTICLE{Ste09,
       author = {{Stewart}, Sarah T. and {Leinhardt}, Zo{\"e} M.},
        title = "{Velocity-Dependent Catastrophic Disruption Criteria for Planetesimals}",
      journal = {\apjl},
         year = 2009,
        month = feb,
       volume = {691},
       number = {2},
        pages = {L133-L137},
          doi = {10.1088/0004-637X/691/2/L133},
       adsurl = {https://ui.adsabs.harvard.edu/abs/2009ApJ...691L.133S}
}

@ARTICLE{Lei12,
       author = {{Leinhardt}, Zo{\"e} M. and {Stewart}, Sarah T.},
        title = "{Collisions between Gravity-dominated Bodies. I. Outcome Regimes and Scaling Laws}",
      journal = {\apj},
         year = 2012,
        month = jan,
       volume = {745},
       number = {1},
          eid = {79},
        pages = {79},
          doi = {10.1088/0004-637X/745/1/79},
archivePrefix = {arXiv},
       eprint = {1106.6084},
 primaryClass = {astro-ph.EP},
       adsurl = {https://ui.adsabs.harvard.edu/abs/2012ApJ...745...79L}
}

@ARTICLE{Bot15,
       author = {{Bottke}, W.~F. and {Vokrouhlick{\'y}}, D. and {Marchi}, S. and {Swindle}, T. and {Scott}, E.~R.~D. and {Weirich}, J.~R. and {Levison}, H.},
        title = "{Dating the Moon-forming impact event with asteroidal meteorites}",
      journal = {Science},
         year = 2015,
        month = apr,
       volume = {348},
       number = {6232},
        pages = {321-323},
          doi = {10.1126/science.aaa0602},
       adsurl = {https://ui.adsabs.harvard.edu/abs/2015Sci...348..321B}
}

@ARTICLE{Hyo17a,
       author = {{Hyodo}, Ryuki and {Genda}, Hidenori and {Charnoz}, S{\'e}bastien and {Rosenblatt}, Pascal},
        title = "{On the Impact Origin of Phobos and Deimos. I. Thermodynamic and Physical Aspects}",
      journal = {\apj},
         year = 2017,
        month = aug,
       volume = {845},
       number = {2},
          eid = {125},
        pages = {125},
          doi = {10.3847/1538-4357/aa81c4},
archivePrefix = {arXiv},
       eprint = {1707.06282},
 primaryClass = {astro-ph.EP},
       adsurl = {https://ui.adsabs.harvard.edu/abs/2017ApJ...845..125H}
}

@ARTICLE{Hyo17b,
       author = {{Hyodo}, Ryuki and {Rosenblatt}, Pascal and {Genda}, Hidenori and {Charnoz}, S{\'e}bastien},
        title = "{On the Impact Origin of Phobos and Deimos. II. True Polar Wander and Disk Evolution}",
      journal = {\apj},
         year = 2017,
        month = dec,
       volume = {851},
       number = {2},
          eid = {122},
        pages = {122},
          doi = {10.3847/1538-4357/aa9984},
archivePrefix = {arXiv},
       eprint = {1711.02334},
 primaryClass = {astro-ph.EP},
       adsurl = {https://ui.adsabs.harvard.edu/abs/2017ApJ...851..122H}
}

@ARTICLE{Hyo17c,
       author = {{Hyodo}, Ryuki and {Charnoz}, S{\'e}bastien and {Ohtsuki}, Keiji and {Genda}, Hidenori},
        title = "{Ring formation around giant planets by tidal disruption of a single passing large Kuiper belt object}",
      journal = {\icarus},
         year = 2017,
        month = jan,
       volume = {282},
        pages = {195-213},
          doi = {10.1016/j.icarus.2016.09.012},
archivePrefix = {arXiv},
       eprint = {1609.02396},
 primaryClass = {astro-ph.EP},
       adsurl = {https://ui.adsabs.harvard.edu/abs/2017Icar..282..195H}
}

@ARTICLE{Hyo17d,
       author = {{Hyodo}, Ryuki and {Charnoz}, S{\'e}bastien},
        title = "{Dynamical Evolution of the Debris Disk after a Satellite Catastrophic Disruption around Saturn}",
      journal = {\aj},
         year = 2017,
        month = jul,
       volume = {154},
       number = {1},
          eid = {34},
        pages = {34},
          doi = {10.3847/1538-3881/aa74c9},
archivePrefix = {arXiv},
       eprint = {1705.07554},
 primaryClass = {astro-ph.EP},
       adsurl = {https://ui.adsabs.harvard.edu/abs/2017AJ....154...34H}
}

@ARTICLE{Hyo18,
       author = {{Hyodo}, Ryuki and {Genda}, Hidenori},
        title = "{Implantation of Martian Materials in the Inner Solar System by a Mega Impact on Mars}",
      journal = {\apjl},
         year = 2018,
        month = apr,
       volume = {856},
       number = {2},
          eid = {L36},
        pages = {L36},
          doi = {10.3847/2041-8213/aab7f0},
archivePrefix = {arXiv},
       eprint = {1803.07196},
 primaryClass = {astro-ph.EP},
       adsurl = {https://ui.adsabs.harvard.edu/abs/2018ApJ...856L..36H}
}

@ARTICLE{Hyo25,
       author = {{Hyodo}, Ryuki and {Genda}, Hidenori and {Madeira}, Gustavo},
        title = "{Pollution resistance of Saturn's ring particles during micrometeoroid impact}",
      journal = {Nature Geoscience},
         year = 2025,
        month = jan,
       volume = {18},
       number = {1},
        pages = {44-49},
          doi = {10.1038/s41561-024-01598-9},
       adsurl = {https://ui.adsabs.harvard.edu/abs/2025NatGe..18...44H}
}

@ARTICLE{Est23,
       author = {{Estrada}, Paul R. and {Durisen}, Richard H.},
        title = "{Constraints on the initial mass, age and lifetime of Saturn's rings from viscous evolutions that include pollution and transport due to micrometeoroid bombardment}",
      journal = {\icarus},
         year = 2023,
        month = aug,
       volume = {400},
          eid = {115296},
        pages = {115296},
          doi = {10.1016/j.icarus.2022.115296},
archivePrefix = {arXiv},
       eprint = {2305.13609},
 primaryClass = {astro-ph.EP},
       adsurl = {https://ui.adsabs.harvard.edu/abs/2023Icar..40015296E}
}

@ARTICLE{Kem23,
       author = {{Kempf}, Sascha and {Altobelli}, Nicolas and {Schmidt}, J{\"u}rgen and {Cuzzi}, Jeffrey N. and {Estrada}, Paul R. and {Srama}, Ralf},
        title = "{Micrometeoroid infall onto Saturn's rings constrains their age to no more than a few hundred million years}",
      journal = {Science Advances},
         year = 2023,
        month = may,
       volume = {9},
       number = {19},
          eid = {eadf8537},
        pages = {eadf8537},
          doi = {10.1126/sciadv.adf8537},
       adsurl = {https://ui.adsabs.harvard.edu/abs/2023SciA....9F8537K}
}

@article{Don91,
  title={A recent cometary origin for Saturn's rings?},
  author={Dones, Luke},
  journal={Icarus},
  volume={92},
  number={2},
  pages={194--203},
  year={1991},
  publisher={Elsevier}
}

@ARTICLE{Cri19,
       author = {{Crida}, Aur{\'e}lien and {Charnoz}, S{\'e}bastien and {Hsu}, Hsiang-Wen and {Dones}, Luke},
        title = "{Are Saturn's rings actually young?}",
      journal = {Nature Astronomy},
         year = 2019,
        month = sep,
       volume = {3},
        pages = {967-970},
          doi = {10.1038/s41550-019-0876-y},
       adsurl = {https://ui.adsabs.harvard.edu/abs/2019NatAs...3..967C}
}

@ARTICLE{Cri25,
       author = {{Crida}, Aur{\'e}lien and {Estrada}, Paul R. and {Nicholson}, Philip D. and {Murray}, Carl D.},
        title = "{The Age and Origin of Saturn's Rings}",
      journal = {\ssr},
         year = 2025,
        month = jul,
       volume = {221},
       number = {5},
          eid = {66},
        pages = {66},
          doi = {10.1007/s11214-025-01189-z},
       adsurl = {https://ui.adsabs.harvard.edu/abs/2025SSRv..221...66C}
}

@ARTICLE{Teo23,
       author = {{Teodoro}, L.~F.~A. and {Kegerreis}, J.~A. and {Estrada}, P.~R. and {{\'C}uk}, M. and {Eke}, V.~R. and {Cuzzi}, J.~N. and {Massey}, R.~J. and {Sandnes}, T.~D.},
        title = "{A Recent Impact Origin of Saturn's Rings and Mid-sized Moons}",
      journal = {\apj},
         year = 2023,
        month = oct,
       volume = {955},
       number = {2},
          eid = {137},
        pages = {137},
          doi = {10.3847/1538-4357/acf4ed},
archivePrefix = {arXiv},
       eprint = {2309.15156},
 primaryClass = {astro-ph.EP},
       adsurl = {https://ui.adsabs.harvard.edu/abs/2023ApJ...955..137T}
}

@article{Cuk16,
  title={Dynamical Evidence for a Late Formation of Saturn's Moons},
  author={{\'{C}}uk, Matija and Dones, Luke and Nesvorn{\'y}, David},
  journal={The Astrophysical Journal},
  volume={820},
  number={2},
  pages={97},
  year={2016},
  publisher={IOP Publishing},
  doi={10.3847/0004-637X/820/2/97}
}

@ARTICLE{Cuk26,
       author = {{{\'C}uk}, Matija and {El Moutamid}, Maryame and {Fuller}, Jim and {Lainey}, Val{\'e}ry},
        title = "{Origin of Hyperion and Saturn's Rings in a Two-stage Saturnian System Instability}",
      journal = {\psj},
         year = 2026,
        month = feb,
       volume = {7},
       number = {2},
          eid = {52},
        pages = {52},
          doi = {10.3847/PSJ/ae422c},
archivePrefix = {arXiv},
       eprint = {2602.09281},
 primaryClass = {astro-ph.EP},
       adsurl = {https://ui.adsabs.harvard.edu/abs/2026PSJ.....7...52C}
}

@article{Ies19,
  title={Measurement and implications of Saturn's gravity field and ring mass},
  author={Iess, L. and Militzer, B. and Kaspi, Y. and Nicholson, P. and Durante, D. and Racioppa, P. and Anabtawi, A. and Galanti, E. and Hubbard, W. and Mariani, M. J. and others},
  journal={Science},
  volume={364},
  number={6445},
  pages={eaat2965},
  year={2019},
  publisher={American Association for the Advancement of Science},
  doi={10.1126/science.aat2965}
}
\bibliographystyle{aasjournalv7}



\end{document}